\documentclass[preprint,12pt]{elsarticle}

 \usepackage{graphicx}

\usepackage{amssymb}
 \usepackage{amsthm} 
 \usepackage{amsmath}

\usepackage{url}
\usepackage{rotating}

\journal{Astroparticle Physics}

\begin{document}

\begin{frontmatter}



\title{Characterization of scatterers for an active focal plane Compton polarimeter}


\author{Sergio Fabiani}

\address{Universit\`a di Roma ``Tor Vergata", Dipartimento di Fisica, Via della Ricerca Scientifica 1, 00133 Rome, Italy \\ INAF-IAPS, Via del Fosso del Cavaliere 100, 00133 Rome, Italy }
\ead{sergio.fabiani@iaps.inaf.it}

\author{Riccardo Campana}
\address{INAF-IAPS, Via del Fosso del Cavaliere 100, 00133 Rome, Italy}

\author{Enrico Costa}
\address{INAF-IAPS, Via del Fosso del Cavaliere 100, 00133 Rome, Italy}

\author{Ettore Del Monte}
\address{INAF-IAPS, Via del Fosso del Cavaliere 100, 00133 Rome, Italy}

\author{Fabio Muleri}
\address{INAF-IAPS, Via del Fosso del Cavaliere 100, 00133 Rome, Italy}

\author{Alda Rubini}
\address{INAF-IAPS, Via del Fosso del Cavaliere 100, 00133 Rome, Italy}

\author{Paolo Soffitta}
\address{INAF-IAPS, Via del Fosso del Cavaliere 100, 00133 Rome, Italy}

\begin{abstract}
In this work we present an active Compton scattering polarimeter as a focal plane instrument able to extend the X-ray polarimetry towards hard X-rays. 

Other authors have already studied various instrument design by means of Monte Carlo simulations, in this work we will show for the first time the experimental measurements of ``tagging efficiency" aimed to evaluate the polarimeter sensitivity as a function of energy. We performed a characterization of different scattering materials by measuring the tagging efficiency that was used as an input to the Monte Carlo simulation. Then we calculated the sensitivity to polarization of a design based on the laboratory set-up. Despite the geometry tested is not optimized for a realistic focal plane instrument, we demonstrated the feasibility of polarimetry with a low energy threshold of 20 keV.
Moreover we evaluated a Minimum Detectable Polarization of 10$\%$ for a 10 mCrab source in 100 ks between 20 and 80 keV in the focal plane of one multilayer optics module of NuSTAR. The configuration used consisted of a doped p-terphenyl scatterer 3 cm long and 0.7 cm of diameter coupled with a 0.2 cm thick LaBr$_{3}$ absorber.

\end{abstract}


\end{frontmatter}



\section{Introduction}
X-ray polarimetry offers the possibility to reach a deep understanding of violent processes taking place around compact objects such as neutron stars, pulsators and black holes which are particularly bright in this energy band.
Focal plane polarimeters based on Compton scattering have been already proposed by \citet{Soffitta2010}, \citet{Hayashida2010} and \citet{Beilicke2011}.
Recent works by \citet{Krawczynski2011a} and \citet{Chattopadhyay2012} evaluated theoretical sensitivities for possible different designs of hard X-ray polarimeters, including focal plane configurations. 
We complement such studies by measuring experimentally the sensitivity of a focal plane active Compton polarimeter. 
We demonstrate a procedure for the instrumental characterization. From the laboratory measurements on the scatterer tagging efficiency, we arrive at the evaluation of the final instrument response by means of Monte Carlo simulations. For such an experiment the realistic sensitivity strongly depends on the lower achievable energy threshold. For example, at 20 keV and 90$^\circ$ of scattering angle, the deposited energy in the scatterer is only 750 eV and the efficiency in detecting the produced faint scintillation signal in organic scatterers needs to be determined. 

In Sect.~\ref{sec:Polarimeter} the design of the focal plane active Compton polarimeter is described. 
In Sect.~\ref{sec:Measurements}  we present the characterization of different scattering materials in terms of tagging efficiency at 22.6 keV with a $^{109}$Cd radioactive source. The best available scatterer sample was also tested at 59.5 keV with an $^{241}$Am source.
In Sect.~\ref{sec:tageffthreshold} the tagging efficiency results are used to evaluate the charge threshold for the signals detection capability in the experimental set-up. Tagging efficiency is evaluated at different energies by applying this threshold to simulated coincidence spectra.
In Sect.~\ref{sec:realpolarimeter} the sensitivity, expressed as Minimum Detectable Polarization (MDP), of a realistic detector design is calculated. Tagging efficiency results are applied to the simulation of a Compton polarimeter based on the experimental set-up. Tested scatterer geometry are not optimized to reach the maximum polarimetric performance, therefore there is still the possibility to improve the sensitivity estimates presented here.

\section{Compton polarimeter description} \label{sec:Polarimeter}

The Compton scattering is effective to measure the polarization in the hard X-ray band. The sensitivity to the polarization of the incident photon is given by the Klein-Nishina cross section \cite{Heitler1954}: 
\begin{equation}
\biggl(\frac{d\sigma}{d\Omega}\biggr)_\mathrm{KN}=\frac{{r_0}^2}{2}\frac{{E^\prime}^2}{{E}^2}\Biggr[ \frac{E}{E^\prime}+\frac{E^\prime}{E}-2\sin^2 \theta \cos^2 \phi \Biggl] \label{eq:KN}
\end{equation}
where 
 \begin{equation}
\frac{E'}{E}=\frac{1}{1+\frac{E}{m_e c^2}(1-\cos \theta)}\label{eq:EsuE}
\end{equation}
$E$ and $E^\prime$ are the energies of the incident and scattered photons respectively, $\theta$ is the scattering angle measured from the incident direction of the incoming photon and $\phi$ is the azimuthal angle measured from the plane containing both the incoming direction and the electric vector of the incident photon. The probability to have a Compton interaction is higher at $\phi=90^\circ$ and $270^\circ$  for a fixed value of $\theta$ and conversely has a minimum for $\phi=0^\circ$ and $180^\circ$.
Thus, linearly polarized photons are preferentially scattered perpendicularly to the direction of the incident photon electric field.
The statistical distribution of the $\phi$ emission directions produced by a beam of polarized radiation is then modulated.
The theoretical modulation factor achievable by an ideal Compton polarimeter is:
\begin{equation}
\mu(\theta)=\frac{N_\mathrm{max}(\theta)-N_\mathrm{min}(\theta)}{N_\mathrm{max}(\theta)+N_\mathrm{min}(\theta)}=\frac{(\frac{d\sigma}{d\Omega})_{\phi=\frac{\pi}{2}}-(\frac{d\sigma}{d\Omega})_{\phi =0}}{(\frac{d\sigma}{d\Omega})_{\phi=\frac{\pi}{2}}+(\frac{d\sigma}{d\Omega})_{\phi =0}}=\frac{\sin^2\theta }{\frac{E}{E^\prime}+\frac{E^\prime}{E}-\sin^2 \theta} \label{eq:Muphi}
\end{equation} 
\begin{figure}
\includegraphics[width=10cm]{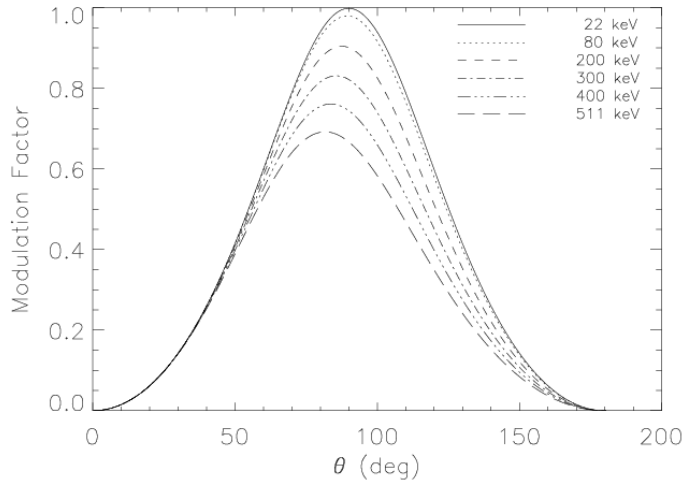}
\caption{Modulation factor dependence on the scattering angle $\theta$ at different energies. Polarized low energy radiation leads to higher modulation factors which maxima occur  for scattering angles approaching 90$^\circ$.}\label{fig:DifferentialMu}
\end{figure}
The modulation factor, as expressed by Eq.~\ref{eq:Muphi}, is shown for different energies in Fig.~\ref{fig:DifferentialMu}.

A good Compton polarimeter should exploit the property of the Klein-Nishina cross section which allows to reach a larger modulation factor for scattering angles $\theta \simeq 90^\circ$ at low energy. 
For a real scatterer the angular distribution of scattered photons is the product of the Klein-Nishina formula and the \textit{scattering function} $S(\chi,Z)$ \cite{Hubbel1975}:
 \begin{equation}
\frac{d\sigma}{d\Omega}=\biggl(\frac{d\sigma}{d\Omega}\biggr)_\mathrm{KN} \cdot S(\chi,Z)\label{eq:KNSF}
\end{equation}
where $Z$ is the atomic number and $\chi=\sin(\frac{\theta}{2})/\lambda[\AA]$ where $\lambda$ is the incident photon wavelength expressed in Angstroms. The scattering function takes into account the influence of the atomic electrons distribution and binding energies. 
A complete review on the scattering function theoretical derivation and numerical calculation is given by \citet{Hubbel1975}. As an example we show the simplified case of the hydrogen atom for which:
 \begin{equation}
S(\chi,H)=1-[F(\chi,H)]^2\label{eq:scat1}
\end{equation}
where $F(\chi,H)=(1+4\pi^2 r_0^2 \chi^2)^{-2}$ is the hydrogen form factor. Therefore the scattering function is:
 \begin{equation}
S(\chi,H)=1-\biggl[\frac{1}{1+4\pi^2 r_0^2 \chi^2}\biggr]^4\label{eq:scat2}
\end{equation}
For large values of the variable $\chi$ the Eq.~\ref{eq:scat2} reduces to 1 which is the hydrogen atomic number.
Substituting the $\chi$ variable in Eq.~\ref{eq:scat2} we have:
 \begin{equation}
S(\chi,H)=1-\biggl[\frac{1}{1+4\pi^2 r_0^2 \bigl(\frac{\sin(\theta/2)}{\lambda}\bigr)^2}\biggr]^4\label{eq:scat3}
\end{equation}
that tells us that the scattering function essentially suppresses forward scattering with respect to the Klein-Nishina formula, since for $\theta=0$ the left term of Eq.~\ref{eq:scat3} vanishes. These properties are verified also for the other elements (see tabulation in \cite{Hubbel1975}).
Therefore the directions of scattering around $\theta=90^\circ$ are still suitable for an efficient detection of the scattered photon. 

The choice of the scattering material should maximize the probability of Compton interaction, while minimizing the photoelectric absorption, which prevails at low energy and sets the low energy threshold of detection. Since the ratio between Compton scattering and photoelectric absorption cross sections is larger for lower $Z$ materials, these ones can be used in the scatterer.
Conversely, a small ratio between scattering and absorption cross sections is required for the absorber. 
Therefore, we consider a polarimeter design in which a low-$Z$ scatterer is coupled with an absorber made of a high-$Z$ material to detect efficiently the scattered photon via the photoelectric effect and thus determining the energy of the incoming photon \cite{Costa1995}. 
Moreover the scatterer should be long enough to ensure a high probability of Compton interaction, 
while maintaining a small width to avoid multiple scattering events. 
 \begin{figure}[htpb] 
 \begin{center}
\begin{tabular}{c}
\includegraphics[scale=0.3]{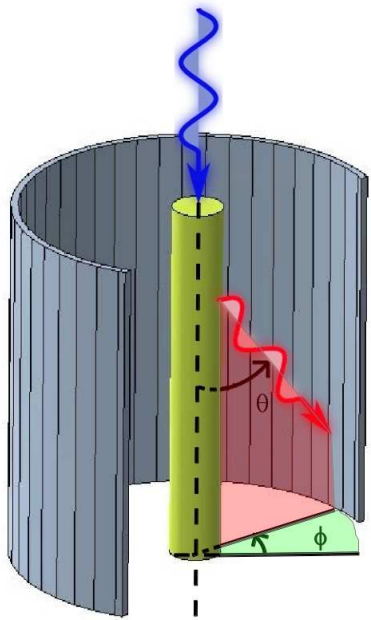}\\
\end{tabular}
\caption{Section of a Compton polarimeter composed by a scattering scintillating rod, surrounded by a cylindrical array of absorber detectors.}\label{fig:GEANT4pol}
 \end{center}
 \end{figure}
Therefore a convenient geometry for a focal plane Compton polarimeter consists of a long and thin scattering scintillating rod surrounded by a cylindrical array of absorber detectors (see Fig.~\ref{fig:GEANT4pol}).
This configuration allows for minimizing systematic effects of spurious modulation, as induced for example by a squared array geometry of absorbers \cite{Hayashida2010,Beilicke2011,Krawczynski2011a}. 
Taking into account only signals in coincidence between the scatterer and the absorbers, it is possible to reduce the background.

 \begin{figure}[htpb] 
 \begin{center}
\begin{tabular}{c}
\includegraphics[scale=0.7]{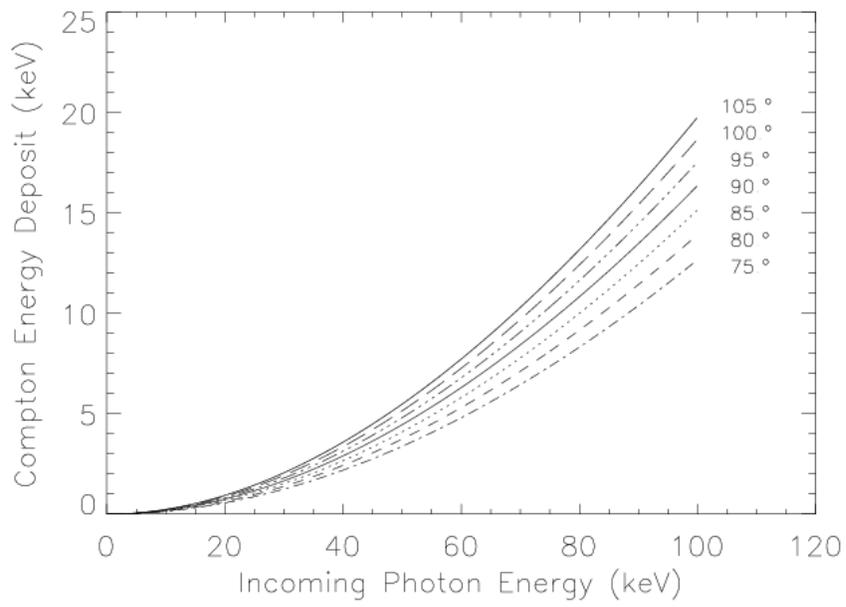}
\end{tabular}
\caption{Compton energy deposits calculated with Eq.~\ref{eq:DeltaE} for some scattering angles around 90$^\circ$.}\label{fig:DeltaE}
 \end{center}
 \end{figure}

From Eq.~\ref{eq:EsuE} the amount of energy released in the scintillator per scattering event is:
\begin{equation}
\Delta E(\theta)= E-E^\prime =\frac{E^2(1-\cos \theta)}{m_e c^2+E(1-\cos \theta)}\label{eq:DeltaE}
\end{equation}
This energy is converted into a scintillation signal. 
Our study is intended to demonstrate the feasibility of a polarimeter able to perform measurements starting from about 20 keV. Such a low energy threshold corresponds to an energy release within the scattering scintillator of about 750 eV for a 90$^\circ$ scattering which is a faint scintillation signal to read-out. In Fig.~\ref{fig:DeltaE} are reported Compton energy deposits calculated with Eq.~\ref{eq:DeltaE} for some scattering angles around 90$^\circ$.

A polarimeter exploiting this design may be placed at the focal plane of a multilayer optics X-ray telescope taking advantage of its large collecting area. It would be possible to perform sensitive polarimetry with a low or negligible background, especially at low energy because of the larger effective area of the optics and the larger intensity of X-ray celestial sources.

\section{Experimental measurements of photon tagging efficiency} \label{sec:Measurements}

The geometry of a polarimeter is the result of a complex trade-off of various design parameters including the maximization of sensitivity, the control of systematics, the energy band of the telescope and the priorities on the astrophysical targets. In our design one of the parameters of highest relevance is the capability to detect weak signals by the scatterer. This determines the efficiency of the system and it is an important driver of the instrument design.
In this work we measure the ``tagging efficiency", which we define as the fraction of events which, after the scattering in the central rod, give rise to a coincidence between the scatterer and the absorber.
To detect a single scattering event in the scintillator we need a scintillation signal at the read-out device higher than a minimum detection threshold.
The characterization of the scatterer material in terms of absolute scintillation output and the choice of the best wrapping to minimize the scintillation signal loss is mandatory. Also the optical coupling between the scintillator and the read-out device is important to collect as much scintillation photons as possible.

\subsection{Scintillators and wrapping materials} \label{subsec:Scatterers}

\begin{table}
\caption{Main properties of the scintillation materials BC-404 and doped p-terphenyl.} 
\begin{tabular}{p{6.5cm}ll}
\hline\noalign{\smallskip} 
  & BC-404 & p-terphenyl  \\
\noalign{\smallskip}\hline\noalign{\smallskip}
Chemical Composition & H$_{11}$C$_{10}$ & C$_{18}$H$_{14}$ \\
Light Output (10$^4$ photons/MeV) & 1.36  & 2.7 \\
Decay time (ns) & 1.8 & 3.7\\
Wavelength of Max Emission (nm) & 408 & 420\\
Refractive index at Wavelength of Max Emission &    1.58     &   1.65   \\
Density (g cm$^{-3}$) &1.023  & 1.23 \\
\noalign{\smallskip}\hline
\end{tabular}\label{tab:scintmat}       
\end{table}

\begin{figure}
\includegraphics[width=12cm]{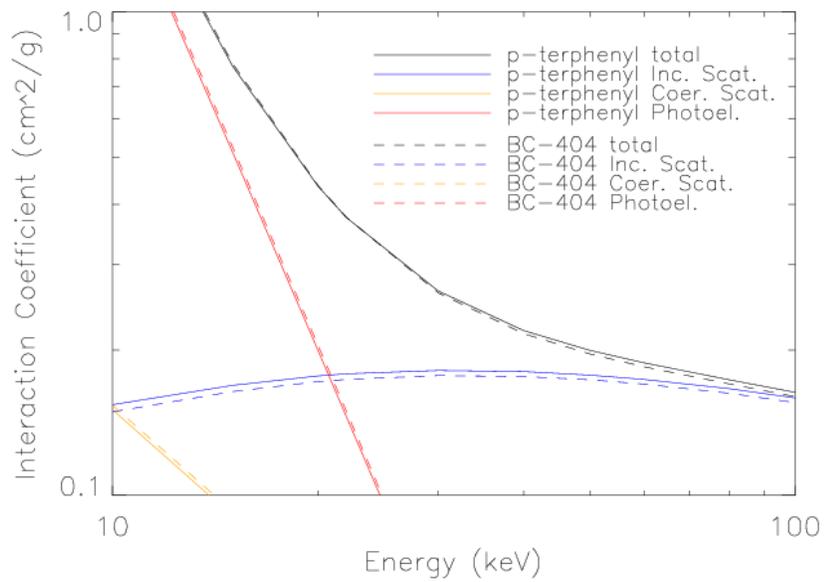}
\caption{P-terphenyl (solid lines) and BC-404 (dashed lines) interaction coefficients. In red is represented the photoelectric absorption, in blue the incoherent scattering, in orange the coherent scattering and in black the total interaction coefficient. P-terphenyl and BC-404 are suitable to study the energy release for Compton scattering down to about 20 keV.  \cite{XCOM}}
\label{fig:PTmu}
\end{figure}

Scintillation materials tested are plastic scintillator BC-404 (Polyvinyltoluene) by Saint-Gobain Crystals \cite{BC404cristal} and doped p-terphenyl crystal by Cryos-Beta \cite{PTCryosBeta} (see Fig.~\ref{fig:PTmu}). Their properties are listed in Tab.~\ref{tab:scintmat}. Doped p-terphenyl has a better scintillation light yield which is useful to achieve a low detection threshold, conversely the BC-404 is characterized by a shorter decay time and it is usually employed for fast coincidences. Since the p-terphenyl crystal has a higher density of about 20$\%$ with respect to the BC-404, it can be used to make a shorter scatterer rod, while maintaining a comparable Compton scattering efficiency.

The wavelengths of maximum emission, 408 nm and 420 nm respectively for BC-404 and doped p-terphenyl, match very well with the sensitivity spectrum of the PMT we employed, a Hamamatsu H10721-110 PMT-sel. \cite{HamaH10721110PMT}, which peaks at about 400 nm and that has a superbialkali photocathode selected by the vendor to have the highest quantum efficiency of the lot, that in our case is 41$\%$.
All scintillators have a cylindrical geometry. Short ones are 1 cm long and have a diameter of 1 cm for both materials. Long ones are 3 cm long and have a diameter of 0.5 cm or 0.7 cm for BC-404 and doped p-terphenyl respectively.

Wrapping materials tested are BC-620 white painting \cite{BC620paint} provided by the manufacturer on a BC-404 sample, commercial PTFE (Teflon), TETRATEX
 tape \cite{TETRATEXref} and a VM2000 double reflecting film. 

\subsection{Measurement procedure}

\begin{figure}
\includegraphics[width=10cm]{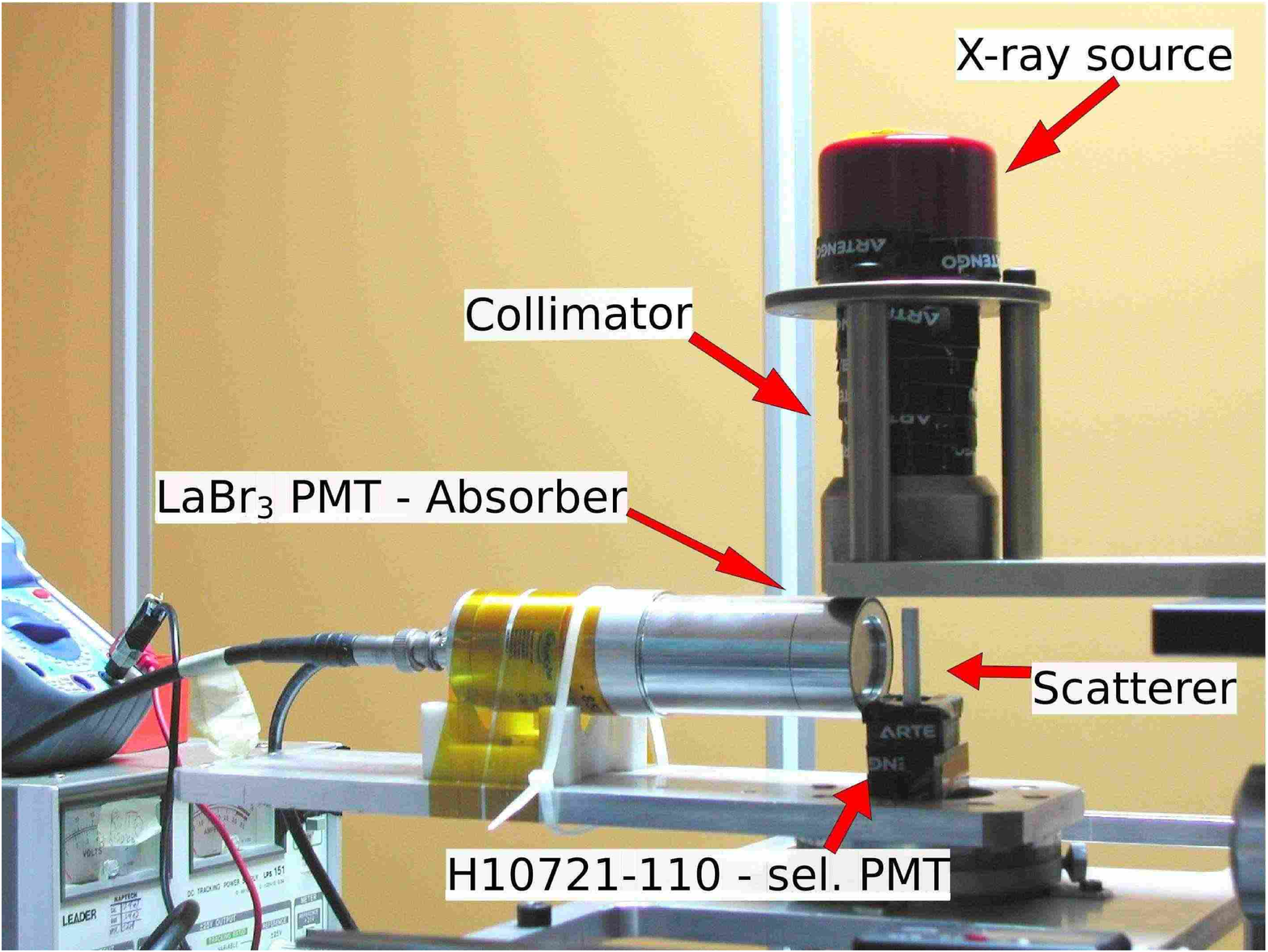}
\caption{Experimental set-up for tagging efficiency measurements. }\label{fig:PhotoSetUp}
\end{figure}

The measurement set-up is shown in Fig.~\ref{fig:PhotoSetUp}.
A radionuclide is held in a lead shielding box. A brass collimator wrapped with lead has a diaphragm on the bottom side to allow the radiation to impinge on the top of the scatterer. Scintillation signal is detected by the Hamamatsu H10721-110 PMT-sel. The sources we used are $^{109}$Cd and $^{241}$Am. $^{109}$Cd emission of interest is given by K$_{\alpha}$ and K$_{\beta}$ superposition of lines of Ag which are not resolved singularly by the scintillator and result in a convoluted line at about 22.6 keV.
$^{241}$Am emission line of interest is at 59.5 keV produced as a consequence of the $\alpha$ decay.
The energy release within the scintillator for Compton scattering at about $\theta \simeq 90^\circ$ for  $^{109}$Cd lines is about 960 eV, and for  $^{241}$Am is about 6.2 keV (see Eq.~\ref{eq:DeltaE} and Fig.\ref{fig:DeltaE}).
The optical contact between the scintillator and the PMT is guaranteed by an optical grease.

Scattered radiation is absorbed by a Brillance 380 Detector manufactured by Saint Gobain Inc. \cite{LaBr3PMT}, which is a sealed PMT with a cylindrical head-on LaBr$_{3}$ crystal 2.5 mm thick and 2.54 cm of diameter behind a 220 $\mu$m thick Beryllium window. 
The whole set-up is covered by a black thick cloth and placed inside a darkroom to minimize the contamination by background stray light. 

\begin{figure}
\includegraphics[width=15cm]{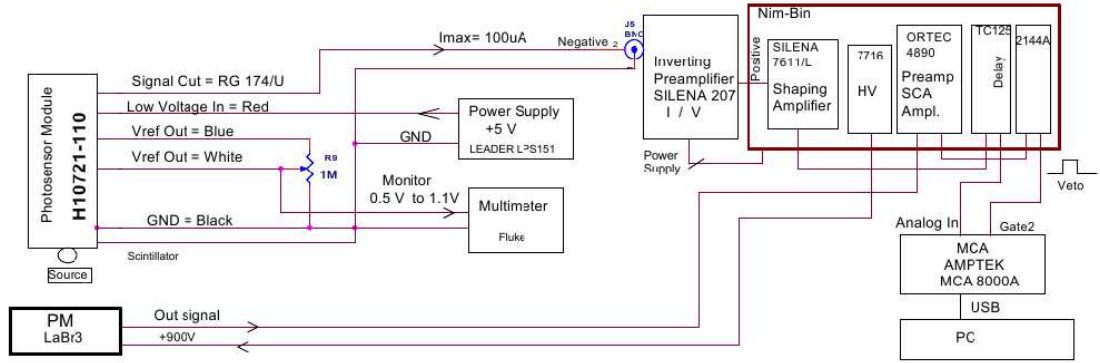}
\caption{Block diagram of the electronic chain employed for the experimental set-up.}\label{fig:ElectricChain}
\end{figure}

The block diagram of the electronic chain is reported in Fig.~\ref{fig:ElectricChain}. The signal from H10721-110 PMT-sel. after pre-amplification (SILENA 207) and amplification (SILENA 7611/L) stages is 1 $\mu$s temporally delayed (TENNELEC TC 215) and sent to a 1024 channels Multi Channel Analyser (MCA Amptek 8000A). The latter provides digitization after applying the \textit{veto}, depending on the signal from the LaBr$_{3}$ PMT. The acquisition from the H10721-110 PMT-sel. is performed only if a photon is detected by the LaBr$_{3}$ PMT within an energy window tuned around the scattered line (ORTEC 4890 Pre-amplifier SCA Amplifier). LaBr$_{3}$ PMT signal triggers a 3 $\mu$s time window (CANBERRA Fast/Slow Coincidence 2144A) to allow MCA to apply the \textit{veto} and to perform digitization.
To minimize dark current rate, a low threshold of 4 ADC channels has been chosen for the MCA acquisition, using an input dynamics range of 0--5 V. Thanks to this instrumental configuration the dynamic at low energy was maximized while excluding dark current from coincidence measurements. We performed also measurements with an input dynamics range 0--10 V, with the same channel threshold, to extend the energy band, to include the $^{109}$Cd peak of photoelectric absorption. The $^{241}$Am photoelectric peak at 59.5 keV was outside the 0--10 V input dynamics range.
ADC spectra obtained with H10721-110 PMT-sel. has been calibrated in term of charge by using a pulse generator for both input dynamics ranges.

\subsection{The scintillator spectrum}

\begin{figure}
\includegraphics[width=10cm]{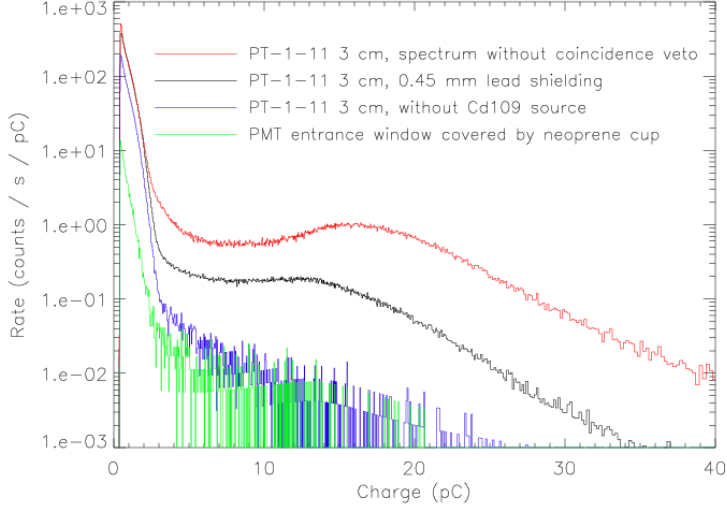}
\caption{Red histogram: spectrum obtained in the PT-1-11 scintillator with $^{109}$Cd source without applying any coincidence veto. On the left of the photoelectric peak a plateau produced by $^{109}$Cd Compton energy deposits of 88.04 keV and by fluorescences generated by the same line exciting the PMT lead cladding. Black histogram: spectrum acquired by blocking direct radionuclide low energy radiation with a lead tape in front of the collimator hole. Blue histogram: spectrum acquired taking away the $^{109}$Cd from the source holder. The Compton plateau disappears, but some low charge signal is still present. Green histogram: the spectrum acquired by replacing the scintillator with a neoprene cap obscuring completely the PMT window.}\label{fig:signalandbackground}
\end{figure}

Spectra obtained with a 3 cm long doped p-terphenyl scintillator illuminated by the $^{109}$Cd source without applying any coincidence veto are shown in Fig.~\ref{fig:signalandbackground}. 
The red histogram represents the spectrum obtained with the set-up of Fig.~\ref{fig:PhotoSetUp}.
It is evident the peak due to photoelectric absorption of $^{109}$Cd emission.
 The black histogram represents the spectrum acquired by blocking the $^{109}$Cd emission with a 0.45 mm thick lead tape in front of the external face of the collimator hole. Around such a tape, the brass diaphragm is still partially transparent to the $^{109}$Cd line at 88.04 keV. 
Therefore the plateau between about 5 and 10 pC, visible in the red and black spectra, is due to Compton energy deposits of $^{109}$Cd 88.04 keV photons passing through the collimator. Moreover such a radiation excites fluorescences at 74.97 keV and 72.08 keV of the lead placed around the Hamamatsu PMT, employed to reduce backscattering. Such fluorescences interact via Compton adding their contribution to the resulting spectrum.
For tagging efficiency measurements the \textit{veto} trigger starts when the LaBr$_{3}$ PMT acquires a signal gated around the Compton scattered line, therefore excluding such a plateau.
The blue histogram represent the spectrum acquired removing the $^{109}$Cd from the source holder. The Compton plateau disappears, but some low charge signal is still present, probably depending on environmental stray light reaching the PMT window or background radiation interacting with the scintillator. 
This interpretation is supported by the fact that using a neoprene cap obscuring completely the PMT window, low amplitude signals are strongly suppressed (green histogram spectrum).
Blue and green histograms demonstrate that the experimental set-up is characterized by a very low background. 

\begin{figure}
\includegraphics[width=10cm]{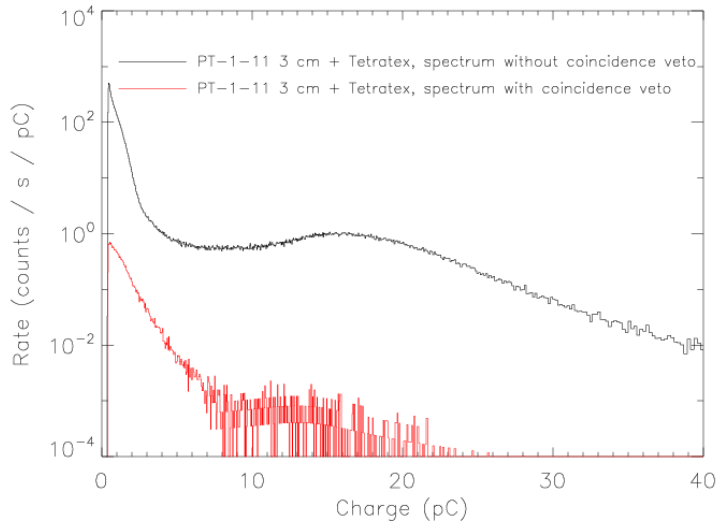}

\caption{$^{109}$Cd Spectra of scintillation signal within the doped p-terphenyl scintillator 3 cm of length and 0.7 cm of diameter wrapped with TETRATEX layer. The spectrum without applying coincidence \textit{veto} is shown in black.
The spectrum obtained by applying coincidence \textit{veto} between the signal from H10721-110 PMT-sel. and LaBr$_{3}$ PMT is shown in red.}\label{fig:coincidenceBC404}
\end{figure}

\subsection{Measurements of tagging efficiency} \label{subsec:Tagging}
In this section we discuss the results of measurement of tagging efficiency obtained with BC-404 and doped p-terphenyl scintillators of various size and wrapped with different materials.

The tagging efficiency is defined as:
 \begin{equation}
\epsilon_\mathrm{tag} = \frac{R_\mathrm{coinc \ net}}{R_\mathrm{tot \ net}} \label{eq:TaggingEff}
\end{equation}
where $R_\mathrm{coinc \ net}$ is the net coincidence rate (without spurious coincidences) between H10721-110 PMT-sel. and LaBr$_{3}$ PMT and $R_\mathrm{tot \ net}$ is the total net rate given only by the scattered events detected in the LaBr$_{3}$ PMT, therefore subtracting the background rate $R_\mathrm{bkg}$. 
The rate of spurious coincidences due to  H10721-110 PMT-sel. electronic noise which accidentally encounters the LaBr$_{3}$ PMT trigger is negligible and it was measured to be (2.3$\pm$0.1)$\cdot$10$^{-3}$ counts s$^{-1}$.  
This value was obtained by shielding the H10721-110 PMT-sel. entrance window with a black cap made of Neoprene below the scattering element for preventing light illumination and applying coincidence \textit{veto}.
Coincidence measurements then allow to fix the problem of dark current (see low charge region of the spectrum represented by the black histogram in Fig.~\ref{fig:coincidenceBC404}).
The background rate $R_\mathrm{bkg}$ detected by the LaBr$_{3}$ PMT is produced by environmental background falling within the energy window of the LaBr$_{3}$ PMT and source photons scattered by the mechanical set-up.  
$R_\mathrm{bkg}$ was measured by removing the scatterer from the above of the H10721-110 PMT-sel. window. 
The background rate in the energy window of $^{109}$Cd scattered emission was acquired three times during long measurements lasting for 3 days each one and it were found to be 0.397$\pm$0.003 counts s$^{-1}$, 0.341$\pm$0.001 counts s$^{-1}$ (MCA dynamic range 0--5V) and 0.331$\pm$0.002 counts s$^{-1}$ (MCA dynamic range 0--10 V). 
For measurements with $^{241}$Am, the background rate was acquired once and it was found to be 0.287$\pm$0.001 counts s$^{-1}$ (MCA dynamic range 0--10 V).
Thus, the Eq.~\ref{eq:TaggingEff}, expressing the tagging efficiency, can be rewritten  as:
\begin{equation}
\epsilon_\mathrm{tag} = \frac{R_\mathrm{coinc}-R_\mathrm{sp \ coinc}}{R_\mathrm{tot}-R_\mathrm{bkg}} \label{eq:TaggingEffExplicit}
\end{equation}
where $R_\mathrm{coinc}$ is the coincidence rate effectively measured, $R_\mathrm{sp \ coinc}$ is the rate of spurious coincidences (negligible in our case) and $R_\mathrm{tot}$ is the total rate effectively measured with the LaBr$_{3}$ PMT.

In some cases multiple evaluations of tagging  efficiency were measured for the same combinations of scintillators and wrapping materials by repeating $R_\mathrm{coinc}$ and $R_\mathrm{tot}$ measurements after adjusting the set-up (i.e. removing and replacing the scatterer) to verify the repeatability of the measurements.
Thus, a weighted average of tagging efficiency value $\overline{\epsilon_\mathrm{tag}}$ have been calculated.
In other cases no multiple measurements were performed so that a single efficiency value is reported.
All results are listed in Tab.~\ref{tab:taggeff}.

\begin{sidewaystable}
\centering
\caption{Tagging efficiencies measured for different scintillator and wrapping materials with $^{109}$Cd and $^{241}$Am. In some case for the same combinations of scintillators and wrapping materials $n$ multiple evaluations of tagging  efficiency were calculated. The acronym PT stands for doped p-terphenyl. In the last column are marked the measurements employed for the further analysis of polarimeter sensitivity.}
\begin{tabular}{lllllllll}
\hline\noalign{\smallskip}
&Scintillator  &Scintillator Dimension &Wrapping & MCA   &number $n$ of & & & Used for \\          
&Material & $h$ (cm) $\times$ $d$ (cm) &	Material&input dynamics (V) &measurements		&$\overline{\epsilon_\mathrm{tag}}$ ($\%$)& $\overline{\sigma_{\epsilon_\mathrm{tag}}}$&Sensitivity\\          
\noalign{\smallskip}\hline\noalign{\smallskip}
\noalign{\smallskip}\hline\noalign{\smallskip}
$^{109}$Cd & &  &	 &  &	&	& &\\
\noalign{\smallskip}\hline\noalign{\smallskip}

&BC-404	&1.0 $\times$ 1.0  &	Teflon& 0-5 &1	&34.44&	0.76&\\				
\\
&BC-404& 	3.0 $\times$ 0.5 & BC-620 paint&0-5 &2  &	14.06&	0.43&\\
& & 	  &+ Teflon& &   &	 &	 &\\
				
&BC-404 &	3.0 $\times$ 0.5 &	Teflon&0-5 &1 &	33.74&	0.80&\\
				
&BC-404 	&3.0 $\times$ 0.5 	&VM2000	&0-5  &1 &36.27&	0.51&\\
\\				
				
&PT-1-11	&3.0 $\times$ 0.7 	&TETRATEX& 0-5 &1	&47.48&	0.23&\\

&PT-1-11 &3.0 $\times$ 0.7 	&TETRATEX&0-10 &1	& 36.15	&0.26&*\\
				
\\

&PT-2-11	&3.0 $\times$ 0.7 	&TETRATEX&0-5 &4	&38.50&	0.16&\\
				
&PT-2-11 &3.0 $\times$ 0.7 	&Teflon&0-5	&1&38.47	&0.39&\\

&PT-2-11 &	3.0 $\times$ 0.7 &	VM2000&0-5 &2	&40.59&	0.16&\\
\\				
&PT-3-11&1.0  $\times$ 1.0 	&Teflon&0-5 &1	&45.5&	1.0&\\
				
&PT-3-11 &1.0  $\times$ 1.0 	&VM2000&0-5	&1&47.20&	0.24&\\
				
&PT-3-11 &1.0  $\times$ 1.0 &	No  wrapping&0-5 &1  &	28.3&	1.9&\\

\noalign{\smallskip}\hline\noalign{\smallskip}
\noalign{\smallskip}\hline\noalign{\smallskip}
$^{241}$Am &  &  &	&  &	& &	&\\
\noalign{\smallskip}\hline\noalign{\smallskip}

&PT-1-11 &3.0 $\times$ 0.7 	&TETRATEX&0-10 &1	&  87.1 &	2.7& *\\

\noalign{\smallskip}\hline
\end{tabular}\label{tab:taggeff}     
 \end{sidewaystable}

Systematic effects were found with respect to the comparison of measurements depending on the rods diameters. The structure holding the PMT glass window has a cavity with a diameter of 1 cm, that encircles the photocathode active region which has a diameter of 0.8 cm.  Scintillator rods 1 cm long have also a diameter of 1 cm. Thus, when coupled with the PMT they are blocked by the structure holding the PMT window, but they exceed the photocathode active region. Even though a fraction of scintillation photons reaching the rod bottom side is lost, such fraction is the same for all 1 cm long rods.
On the contrary, rods which are 3 cm long have a diameter of 0.7 cm (doped p-terphenyl) and 0.5 cm (BC-404), therefore smaller than the PMT active window. Moreover, they can move partially out from the PMT active window causing some scintillation photons loss. Such effect can affect the reproducibility of measurements, allowing only the evaluation of maximum values for tagging efficiency. Therefore it make sense to consider multiple measurements for different wrapping/scintillator combinations looking for the reproducibility of results. Whenever single measurements values were compatible within the $3\sigma$ error we performed the average of tagging efficiency to obtain results reported in Tab.~\ref{tab:taggeff}.
Therefore, due to the diameter size larger than the active window of the PMT for 1 cm long rod, we cannot compare directly collected scintillation signal from rods of different length. Despite this fact we notice that the length is an important parameter: longer rods increase scattering efficiency while leading to a larger scintillation signal loss, which must be reduced using an efficient wrapping material.

Two rods of doped p-terphenyl 3 cm long with 0.7 cm of diameter were tested. One of them (PT-1-11) shows a higher tagging efficiency with respect to the other (PT-2-11). For preserving the better sample, we decided to leave it wrapped with its original TETRATEX layer as delivered by the vendor. 
The difference of tagging efficiency arises from a larger light output of the PT-1-11 with respect to PT-2-11 allowing to get more coincidences and therefore to achieve a higher tagging efficiency.  At a visual inspection crystals do not  show any evident difference, but even small structural defects may contribute to change scintillation photons production and collection properties.

\subsection{Comparison between doped P-terphenyl and BC-404}\label{subsec:materials}
\begin{figure}
\includegraphics[width=10cm]{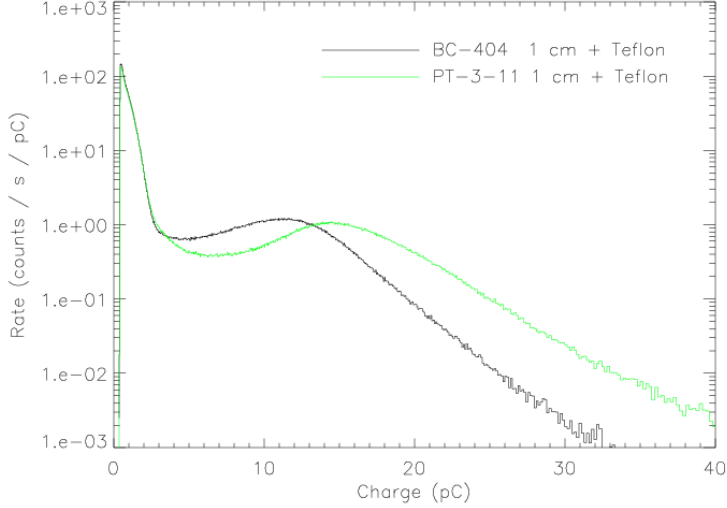}
\caption{Spectra of doped p-terphenyl PT-3-11 (green) and BC-404 (black) 1 cm long rods both wrapped with Teflon. A stronger scintillation signal  is achieved by employing PT-3-11 which has a theoretical light output about 2 times larger than BC-404 (see Tab.~\ref{tab:scintmat}).}\label{fig:PT311vsBC404}
\end{figure}

Doped p-terphenyl is characterized by an expected larger light output with respect to BC-404 (see Tab.~\ref{tab:scintmat}). This is confirmed by the spectra reported in Fig.~\ref{fig:PT311vsBC404} showing the comparison of 1 cm long scintillators wrapped with Teflon tape.  
Therefore, the photoelectric absorption peak of doped p-terphenyl scintillator is shifted towards a larger value of charge collection if compared with BC-404.
Since doped p-terphenyl is able to produce more scintillation photons with respect to BC-404, more events will be detectable and therefore doped p-terphenyl tagging efficiency is expected to be larger than BC-404 one. 
This fact is confirmed by results reported in Tab.~\ref{tab:taggeff}, where doped p-terphenyl scintillators are systematically more efficient than BC-404 ones. In particular tagging efficiencies of rods of 1 cm of length, wrapped with Teflon, are $(34.44\pm0.76) \%$ for BC-404 and $(45.5\pm 1.0) \%$ for PT-3-11. Therefore BC-404 tagging efficiency is $75.7\%$ of the p-terphenyl tagging efficiency. If rods of 3 cm of length, wrapped with Teflon,  are considered the difference in tagging efficiency decreases. In fact the BC-404 tagging efficiency is $(33.74\pm0.80) \%$ while the PT-2-11 tagging efficiency is $(38.47\pm0.39) \%$. Therefore the BC-404 tagging efficiency is $87.7\%$ of the p-terphenyl one.
This demonstrate how the p-terphenyl crystal is less efficient to collect scintillation photons, since if longer rods are considered its tagging efficiency decreases more with respect to the BC-404.
The p-terphenyl crystal could present internal inhomogeneities which reduce its scintillating and visible light collection performance. This problem can be more significant for long rods with respect to shorter one.

\subsection{Comparison between VM2000, TETRATEX and Teflon}
A good wrapping plays a crucial role to enhance the tagging efficiency owing to the preservation of scintillation signal.
\begin{figure}
\includegraphics[width=10cm]{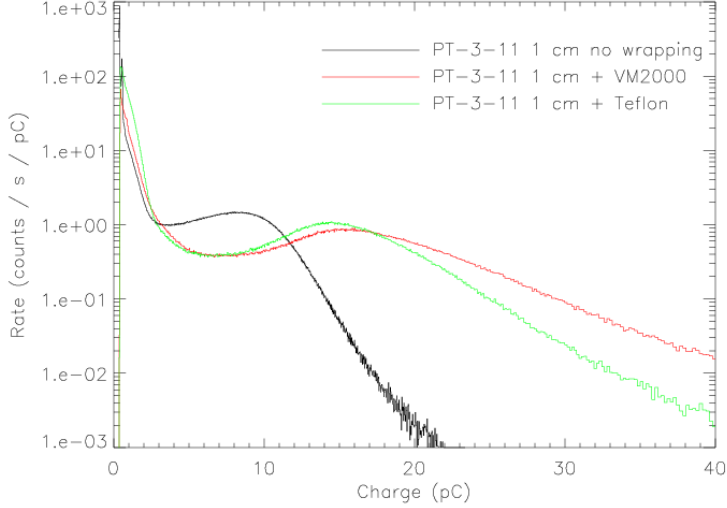}
\caption{Spectra from doped p-terphenyl PT-3-11 sample (1 cm long rod) wrapped with VM2000 (red), Teflon (green) and without any wrapping (black). The VM2000 reflecting foil allows to preserve the larger fraction of scintillation signal, and it is also the one that leads to the highest tagging efficiency.}\label{fig:betterWrappingPT}
\end{figure}
The fact that a better tagging efficiency is associated to a higher collection of light can be demonstrated by comparing the peaks of photoelectric absorption from doped p-terphenyl  1 cm long rods  wrapped with different materials (see Fig.~\ref{fig:betterWrappingPT}) and tagging efficiency results. VM2000, which allows for collecting the most intense scintillation signal, it is also the one which leads to the highest tagging efficiency.
Tab.~\ref{tab:taggeff} shows clearly as VM2000 reflecting layer allows to obtain a higher tagging efficiency with respect to the other wrapping materials.

Here we omit any discussion on the other aspects such as the uniformity of the wrapping and the impact on the systematics. These considerations may be faced for a real polarimetric experiment, but they are not relevant to the present discussion.

\section{The tagging efficiency as a low amplitude signal threshold}\label{sec:tageffthreshold}

The energy deposited by a scattered photon produces a scintillation signal which is larger for higher energy release (see Eq.~\ref{eq:DeltaE}) and for smaller scintillation light loss. Since the tagging efficiency identifies the fraction of such a Compton energy deposit effectively read-out above the detection threshold, which is independent on the incident radiation energy, different values of tagging efficiency are expected. The higher is the energy of incident radiation, the higher will be the tagging efficiency.

To evaluate the sensitivity of a Compton polarimeter across the X-ray energy band, it is necessary to quantify the tagging efficiency energy dependence across the energy spectrum. A way to obtain such an information is to perform direct experimental measurements of tagging efficiency for all the energies of interest.
Unfortunately there is a limited number of long lived radioactive sources emitting lines at the energies of our interest, not emitting also higher energy lines interfering with the measurements.
An alternative strategy is necessary. We measured the tagging efficiency of $^{109}$Cd and $^{241}$Am (see Sect.~\ref{sec:Measurements}). Then, by simulating the coincidence Compton spectra of $^{109}$Cd and $^{241}$Am with the experimental set-up shown in Fig.~\ref{fig:PhotoSetUp}, we found the threshold corresponding to the measured tagging efficiency for each radionuclide spectrum. The two values found are two estimates of the same threshold which is independent on the incident energy, therefore we calculated their weighted average.
Subsequently, we simulated coincidence Compton spectra for different incident energies with the same set-up, and we calculated the tagging efficiency corresponding to the threshold we obtained before. Finally, by simulating a realistic polarimeter design (see Fig.~\ref{fig:GEANT4pol}) and multiplying such tagging efficiencies for the Compton polarimeter efficiencies at the same energies, it is possible to obtain the total efficiency and then to evaluate the sensitivity for the polarimeter design considered.

 \subsection{Simulations of laboratory set-up}

A simulator of the the Compton energy deposition within the scintillator in the experimental set-up (see Fig.~\ref{fig:PhotoSetUp}) was developed using the GEANT 4 toolkit \cite{Agostinelli2003}.
The simulated scintillator chosen is the doped p-terphenyl PT-1-11 wrapped with TETRATEX to reproduce measures performed with the input dynamics range 0--10 V with $^{109}$Cd and $^{241}$Am (see measures labelled with ``*" in Tab.~\ref{tab:taggeff}).
The simulator output file reports the energy deposits occurred within the scintillator at different depths. Coincidences between scatterer and absorber are tagged with a specific flag to be later analysed.

The GEANT 4 simulator only models the energy losses in the scintillator.
We choose to model the scintillation light propagation within the scatterer by means of a Monte Carlo simulator specifically developed for this purpose.
The input of this simulation stage are the coincident events energy deposits and their interaction depth within the scintillator as obtained by the GEANT 4 simulator. The output are the coincidence spectra of $^{109}$Cd and $^{241}$Am suitable to be compared with the experimental ones as obtained by the H10721-110 PMT-sel.

In this Monte Carlo simulator the X-ray photons are assumed to be Compton scattered along the scintillator vertical axis. Since the X-ray beam is collimated and centred, this approximation is good.
The number of scintillation photons produced for each X-ray scattering event is extracted from a Poissonian distribution of expected value:
 \begin{equation}
m_{0}=\frac{\Delta E}{\delta}\label{eq:m0}
\end{equation}
where $\Delta E$ is the energy deposit as given by GEANT 4 simulator and $\delta$ is the scintillator light output (see Tab.~\ref{tab:scintmat}).

The path length $s$ of each scintillation photon is evaluated extracting a random value from an exponential distribution that depends on the mean free path $\lambda$ of visible light within the scintillation material. Therefore $s$ is given by:
 \begin{equation}
s=-\lambda \ln(R)\label{eq:scintpathlenght}
\end{equation}
where $R$ is a random number between 0 and 1. The mean free path value that gives the better agreement with measured data is 2.0 cm. Indeed the scintillation light extinction at small distance (few centimetres) from the visible light production site is large \cite{Nicoll1970} and $\lambda$ is very small.
The probability of scintillation light to be reflected or transmitted through a discontinuity layer between two different materials is evaluated by means of Fresnel equations. Also the total internal reflection, if conditions are verified, is assumed. Since the scintillator is wrapped with the TETRATEX diffuser on the top and side faces, visible light passing through its edges encounters air and then it is diffused by TETRATEX. Therefore in this case the discontinuity between p-terphenyl and air is taken into account to evaluate the probability of reflection or diffusion.
Light exiting from the scintillator faces enters into the diffuser which has a probability of 95$\%$ \cite{TERTATEXreflectivity} to diffuse it back. Since the diffuser produces a randomisation of directions the diffusion angle is a random value between $-\pi$ and $\pi$ with respect to the normal to the discontinuity layer. 
In the Monte Carlo simulator we developed, scintillation photons propagate within vertical planes passing thought the scatterer vertical axis, therefore reflection by the scintillator edges, but also diffusion by the wrapping, does not produce any variation of the azimuthal direction.
When a scintillation photon reaches the bottom face of the scatterer the following discontinuities are considered: 
\begin{enumerate}
\item between the scintillator and the optical grease, 
\item between the optical grease and the PMT glass window,
\item between the PMT glass window and the photocathode
\end{enumerate}
The refractive indexes are: for the p-terphenyl scintillator 1.65 (see Tab.~\ref{tab:scintmat}), for the optical grease 1.5, for the PMT glass window 1.55 and for the superbialkali photocathode 2.15 (the real part) and 1.2 (the imaginary part). This last value was estimated from \citet{Motta2005} that studied the reflection properties of bialkali photocathodes.
Then survived photons crossing the PMT glass window are converted into photoelectrons by the photocathode and start the multiplication avalanche.

The average number of photoelectrons generated by the photocathode is $\mu=mq$ where $q$ is the photocathode quantum efficiency. 
Assuming all photoelectrons are collected by the first dynode, the probability that $n$ photoelectrons will start the multiplication is given by the Poisson relation \cite{Bellamy1994}:
\begin{equation}
P(n; \mu)=\frac{\mu^n e^{-\mu}}{n!}\label{eq:poisson}
\end{equation}

The multiplication process for $n$ photoelectrons, each starting a mutually independent chain, can be approximated by a Gaussian distribution, if the coefficient of secondary electron emission by the first dynode is large ($>$ 4) and the coefficient of secondary electron collection by the initial dynodes multiplication stages is close to one \cite{Bellamy1994}:
\begin{equation}
G_n(x)=\frac{1}{\sigma_1 \sqrt{2 \pi n}}e^{-\frac{(x-nQ_1)^2}{2 n{\sigma_1}^2}}\label{eq:gauss}
\end{equation} 
where $x$ is the PMT output signal expressed in charge, $Q_1$ is the PMT charge output signal at the end of the multiplication chain for one photoelectron, $\sigma_1$ is the standard deviation of the one photoelectron peak charge distribution.
Obviously $Q_1= e g$  where $e$ is the elementary charge and $g$ is the PMT gain. 
For $n=0$ we have simply zero charge at the output signal.
The convolution of Poisson (Eq.~\ref{eq:poisson}) and Gaussian distributions (Eq.~\ref{eq:gauss}) gives \cite{Bellamy1994}:
 \begin{equation}
S_\mathrm{output}(x)=P(n;\mu) \otimes G_n(x)=\sum_{n=1}^{n_\mathrm{max}}{\frac{\mu^n e^{-\mu}}{n!} \frac{1}{\sigma_1 \sqrt{2 \pi n}}e^{-\frac{(x-nQ_1)^2}{2 n{\sigma_1}^2}} }\label{eq:poisgaussconvolution}
\end{equation}
The Eq. \ref{eq:poisgaussconvolution} describes the PMT response to an event which produces $\mu$ average photoelectrons at the photocathode. 

In the case of the $^{109}$Cd source, most of the PMT output is given by low amplitude signals since the energy deposit within the scintillator is small and the scintillation light transport will allow only to a small number of scintillation photons to reach the PMT entrance window.
The experimental characterization of small amplitude signals is an important issue in order to model properly the photomultiplier response.
A weak signal like the pulse from a single primary electron can show significant fluctuations. In the spectrum of single photoelectron a fraction of 10--20$\%$ could be due to small pulses below 1/3 of peak amplitude position. Those signals are produced by photoelectrons that are inelastically back-scattered by the first dynodes \cite{ManualPMT, Knoll2010}. They distort the peak shape producing a wing in the low amplitude part of the spectrum. Therefore the single photoelectron peak ($n=1$ in Eq.~\ref{eq:poisgaussconvolution}) is an intrinsic characteristic of the specific PMT. 
The measurement of the single photoelectron spectrum requires a proper experimental set-up. For example a single photon light pulse from a LED can be used to trigger the PMT acquisition \cite{Barnhill2008}. 
At the moment we do not have the possibility to perform such a kind of study to measure properly the single photoelectron spectrum. 
There is an other method in the literature \cite{Bauleo2004} according to which the single photoelectron spectrum is measured by integrating the dark current after  the PMT was left for a long time in the dark. However in this case the signal produced by single electrons emitted for thermal agitation from the photocathode can be distorted by low amplitude signals coming from the emission of electrons from subsequent dynodes. 
Therefore, we did not perform the characterization of the single photoelectron spectrum and we decided to modify Eq.~\ref{eq:gauss} and then Eq.~\ref{eq:poisgaussconvolution} to reproduce low amplitude signals at least by maintaining the total probability of the Gaussian statistics. The low amplitude wing of the Gaussian of Eq.~\ref{eq:poisgaussconvolution}, when $n$ is a small number, forms a not negligible part of the spectrum and therefore of the Gaussian probability. We mirrored the Gaussian lower tail for negative values of the the charge. Following this procedure the total probability of the Gaussian is preserved.

\subsection{The energy dependent tagging efficiency as derived by the detection threshold in $^{109}$Cd and $^{241}$Am} 

\begin{figure}
\includegraphics[width=8cm]{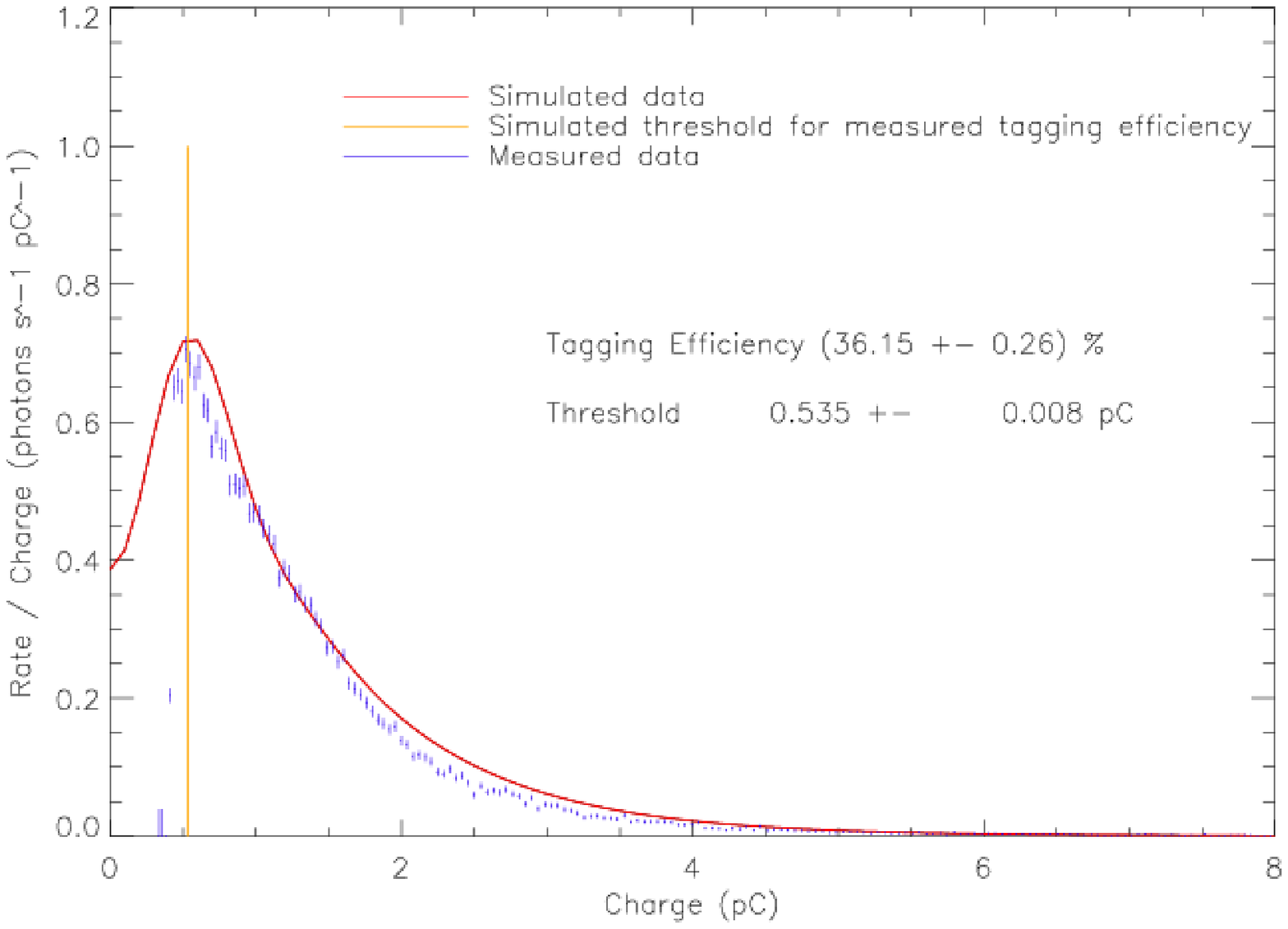} 
\includegraphics[width=8cm]{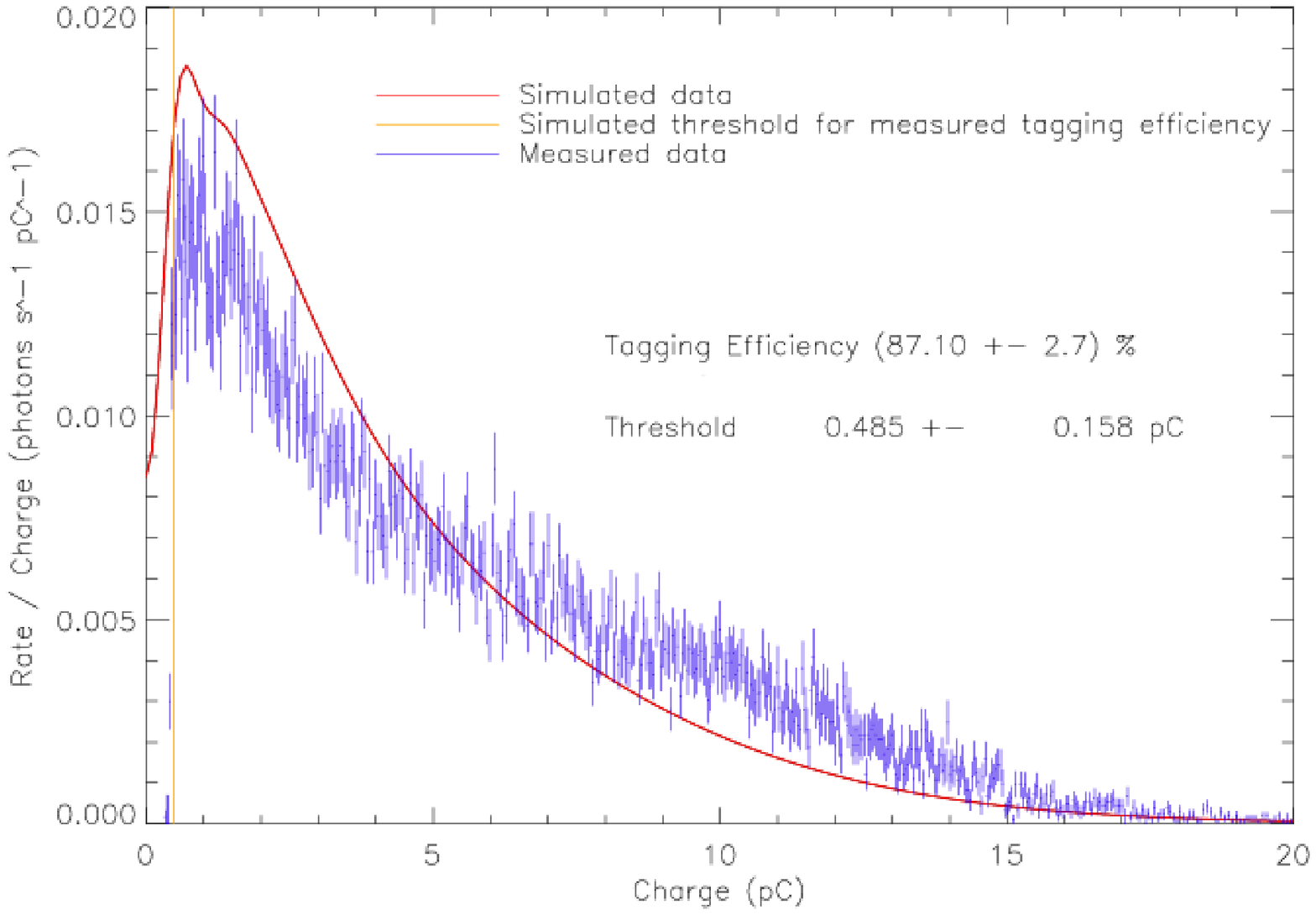}
\caption{$^{109}$Cd (superposition of lines at about 22.6 keV)and $^{241}$Am (59.5 keV) measured (blue) and simulated (red solid line) coincidence spectra. The orange line is the charge threshold obtained from simulated spectra applying the measured tagging efficiency.}\label{fig:simCd109Am241}
\end{figure}

Eq.~\ref{eq:poisgaussconvolution} is used to simulate coincidence spectra of $^{109}$Cd (superposition of lines at about 22.6 keV) and $^{241}$Am (59.5 keV) (see Fig.~\ref{fig:simCd109Am241}) whose measured tagging efficiency is reported in Tab.~\ref{tab:taggeff} (measurements labelled with ``*"). Since the measured tagging efficiency is known, the corresponding detection threshold $x_\mathrm{thr}$, expressed in charge, is the one that gives the same simulated tagging efficiency calculated as follows:
\begin{equation}
\epsilon_\mathrm{tag}(x_\mathrm{thr})=\frac{\Biggl[  \biggl( N_\mathrm{tot}-\sum_{i=0}^{N_\mathrm{tot}}{P(n=0; \mu_i)} \biggr) \cdot \mathcal{I}(S_\mathrm{output}(x_\mathrm{thr})) \Biggr]}{N_\mathrm{tot}} \label{eq:taggingeffsim}
\end{equation}
where
 \begin{equation}
\mathcal{I}(S_\mathrm{output}(x_\mathrm{thr})) =\frac{\int_{x_\mathrm{thr}}^{x_\mathrm{max}}{S_\mathrm{output}(x)}}{  \int_{0}^{x_\mathrm{max}}{S_\mathrm{output}(x)}   }dx\label{eq:I}
\end{equation}
In Eq.~\ref{eq:taggingeffsim}  $N_\mathrm{tot}$ is the total number of simulated X-ray coincidence scattering events simulated in coincidence, the term $\sum_{i=0}^{N_\mathrm{tot}}{P(n=0; \mu_i)}$ gives the total number of events which extract zero photoelectrons from the photocathode. Thus, $\biggl(N_\mathrm{tot}-\sum_{i=0}^{N_\mathrm{tot}}{P(n=0; \mu_i)}\biggr)$ gives the number of events that produce an output charge signal at the end of dynodes chain (which have $n \ge 1$). The term $\mathcal{I}(S_\mathrm{output}(x_\mathrm{thr}))$ represents the fraction of simulated charge signal above the charge threshold. The weighted average of $^{109}$Cd and $^{241}$Am charge threshold is finally evaluated.

Measured (blue points) and simulated spectra (red solid line) are reported in Fig.~\ref{fig:simCd109Am241}. 
The simulated detection threshold $x_\mathrm{thr}$ (vertical orange solid line) is also shown in Fig.~\ref{fig:simCd109Am241}. It can be compared with the hardware experimental threshold corresponding to the sharp cut of the spectra on the left.

Tagging efficiencies for 20, 25, 30, 35, 40, 45, 50, 60, 70, 80, 90 and 100 keV are obtained by simulating each coincidence spectrum and substituting into the Eq.~\ref{eq:taggingeffsim} the threshold value $x_\mathrm{thr}$ previously calculated. 
\begin{figure}
\includegraphics[width=12cm]{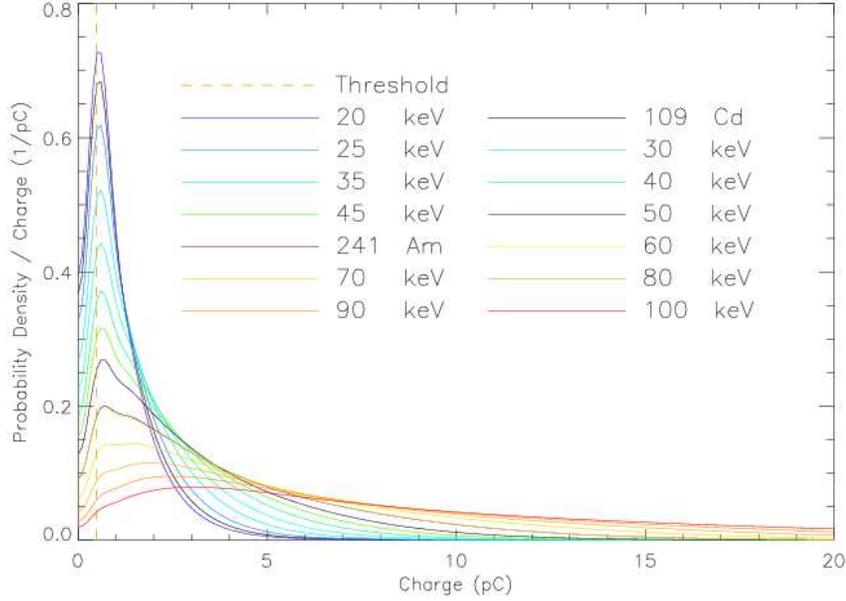} 
\caption{Comparison of coincidence simulated spectra at different energies.}\label{fig:SimChargeSpectrumConfrontation}
\end{figure}
Coincidence spectra at different energies are compared in Fig.~\ref{fig:SimChargeSpectrumConfrontation}.

\begin{figure}
\includegraphics[width=12cm]{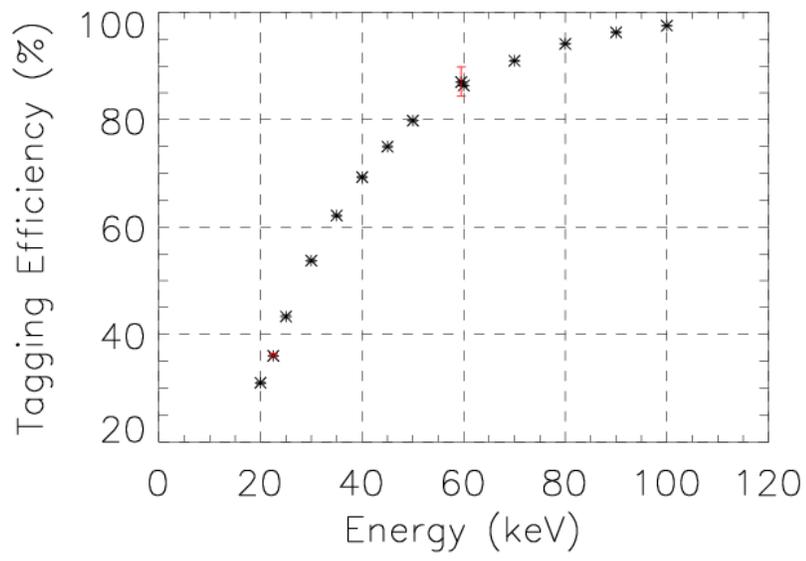} 
\caption{Simulated tagging efficiency (black) plotted with measured one (red) for $^{109}$Cd and $^{241}$Am.}\label{fig:taggeffplot}
\end{figure}

Tagging efficiency values obtained are reported in Fig.~\ref{fig:taggeffplot}. The efficiency increases from lower to higher energies and experimental measured values are compatible with simulated ones. 

\section{From tagging efficiency to realistic polarimeter sensitivity}\label{sec:realpolarimeter}

In this section we show the simulated sensitivity of a polarimeter design based on the experimental set-up characterized in previous sections.

Polarimeters performance can be evaluated by means of the Minimum Detectable Polarization (MDP), that is the minimum polarization degree detectable at a certain confidence level. \citet{Weisskopf2010} calculated the MDP for the 99$\%$ confidence level as:
 \begin{equation}
MDP(99\%)=\frac{4.29}{\mu R_\mathrm{source} }\sqrt{\frac{B+R_\mathrm{source}}{ T}}\label{eq:MDP}
  \end{equation}  
where $R_\mathrm{source}$  is the coincidence rate from the source,  $B$ is the background coincidence rate, $T$  is the observing time and $\mu$ is the modulation factor. 

In case of a focal plane polarimeter, for which the background can be neglected because the instrument is largely source dominated (possibly by the addition of an external anticoincidence detector), the MDP formula simplifies in:
 \begin{equation}
MDP(99\%)\simeq \frac{4.29}{\mu \sqrt{T R_\mathrm{source}}} \  \   \  \ \ \ \mbox{if}  \  \   \  \ \ \ B \ll R_\mathrm{source} \label{eq:MDPsimple}
  \end{equation}  

The coincidence source rate detected is proportional to the tagging efficiency $\epsilon_\mathrm{tag}$, such that Eq.~\ref{eq:MDPsimple} gives:
\begin{equation}
MDP \propto \frac{1}{\sqrt{R_\mathrm{source}}} \propto \frac{1}{\sqrt{\epsilon_\mathrm{tag}}} \label{eq:MDPpropto}
\end{equation}

A polarimeter design as described in Sect.~\ref{sec:Polarimeter} was simulated by means of the GEANT 4 toolkit.
The scatterer is a 3 cm long rod which diameter is 0.7 cm, made of doped p-terphenyl (the material which gave the higher light output response) and wrapped with TETRATEX. A beam of monochromatic radiation impinges on axis at the center of the scatterer upper face. A LaBr$_{3}$ absorber is placed all around the scatterer at a distance of 5 cm away from the scatterer axis. The distance between the scatterer and the absorber is an important geometric parameter. If they are placed at a short distance scattering angles quite different from 90$^\circ$ are possible for coincidences and then a low modulation is achieved (see Fig.~\ref{fig:DifferentialMu}).
The absorber is subdivided in 48 pixels, 3 cm long and 0.2 cm thick, laying parallel to the rod extension to measure the azimuthal directions of scattering. We derived from this simulation the efficiency as a function of energy, that we then multiplied by the tagging efficiency (evaluated in Sect.~\ref{sec:tageffthreshold}), and the modulation factor.

To evaluate the polarimeter performance then we calculated the MDP achievable at the focal plane of one optics module of NuSTAR \cite{NuSTAROptics, Harrison2010}, that will observe for the first time the sky in the hard X-rays between 5 and 80 keV. NuSTAR telescope is composed by two 10.14 m focal length identical optics modules based on Wolter I geometry employing multilayer technology. 

\begin{figure}
\includegraphics[scale=0.45, angle=90]{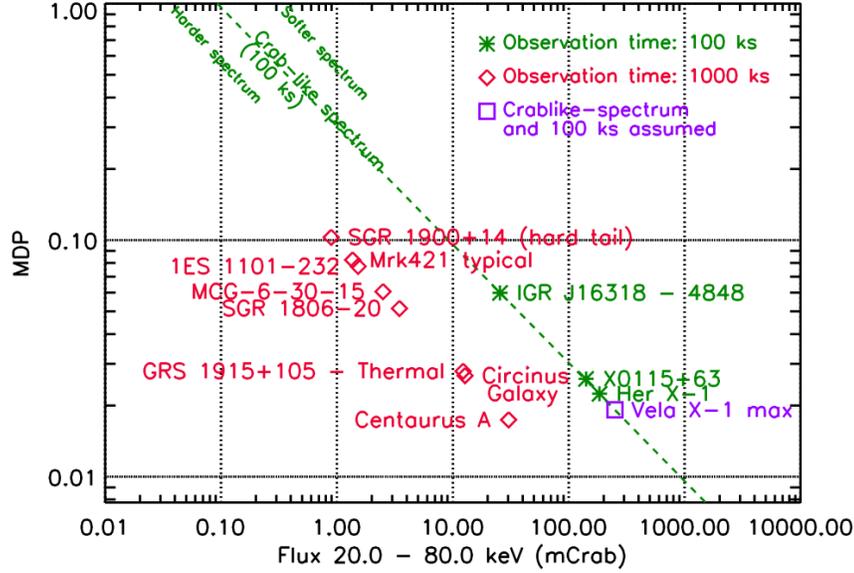} 
\caption{MDP for a Compton polarimeter based on the experimental set-up at the focal plane of one module of NuSTAR telescope. For evaluating the MDP achievable with both optics modules the plotted values must be divided by a factor $\sqrt{2}$, because the MDP scales with the root square of the effective collecting area. 
}\label{fig:MDPplot}
\end{figure}

In Fig.~\ref{fig:MDPplot} is reported the MDP for various sources. An MDP of about 10$\%$ between 20 and 80 keV is obtained for 10 mCrab sources in 100 ks of observation. The MDP achievable with both optics modules can be evaluated taking into account that the coincidence source rate $R_\mathrm{source}$ scales with the effective collecting area $A_\mathrm{eff}$, so that from Eq.~\ref{eq:MDPpropto} we have:
 \begin{equation}
MDP \propto \frac{1}{\sqrt{R_\mathrm{source}}} \propto \frac{1}{\sqrt{A_{\mathrm{eff}}}}\label{eq:MDPAeff}
\end{equation}
Thus, doubling the area (i.e. observing with both optics modules) the MDP reduces by a factor $\sqrt{2}$.

Neutron stars with cyclotron lines could be observed, if the lines fall within the 20~--~80~keV energy band. Among them an MDP lower than 3$\%$ is achievable for Vela~X-1 which shows cyclotron lines at about 23 and 50 keV  \cite{Kreykenbohm2002}, X0115+63 which presents multiple cyclotron absorption features from 10 to 50 keV \cite{Santangelo1999, Heindl2000} and Her~X-1 which has a line at about 40~keV \cite{Truemper1978, Gruber2001, Vasco2011}.

From the point of view of the optimization of the polarimeter we can make some considerations.
Since a long crystal scattering rod could have internal inhomogeneities that reduce the collection of scintillating light (see Sect.~\ref{subsec:materials}), the choice to enhance the detector sensitivity by means of enhancing the Compton efficiency with a rod longer then the 3 cm, requires probably to change the scintillating material passing to a plastic one (i.e. BC-404).
For what concerns the wrapping material, the VM2000 reflecting film is the one which allows to obtain the higher tagging efficiency. For our experimental measurements we decided to preserve the most efficient crystal sample by avoiding to replace its TETRATEX wrapping with the VM2000 film, but a smaller MDP would be expected in this other a case.

\section{Conclusions}

We defined a procedure based on experimental laboratory measurements and Monte Carlo simulations aimed to characterize the physical response of an active Compton polarimeter composed by a central low-$Z$ scatterer rod surrounded by a high-$Z$ absorber. 
The cylindrical geometry of the absorbers array allows for reducing systematics effect. 
In this study the experimental characterization is employed to fix simulator parameters needed to evaluate the instrument response. 
We characterized BC-404 (Polyvinyltoluene) plastic scintillator and doped p-terphenyl crystal. The latter showed a larger light output leading to a higher tagging efficiency with respect to the other one. We verified also that VM2000 reflecting film is the best wrapping material with respect to PTFE (commercial Teflon), TETRATEX and BC-620 white painting, giving a higher tagging efficiency depending on a better preservation of scintillation signal within the scintillator.

We concluded by showing the sensitivity of a polarimeter design based on the laboratory set-up with a 3 cm long doped p-terphenyl rod wrapped with TETRATEX coupled with LaBr$_{3}$ absorber placed at the focal plane of one of the two optics module of NuSTAR telescope.
The Minimum Detectable Polarization achievable is 10$\%$ between 20 and 80 keV for a 10 mCrab source in 100 ks of observation.
The sensitivity of this kind of polarimeter can be improved. If a longer scatterer is employed, the Compton efficiency is enhanced. However a longer p-terphenyl crystal could be affected by internal inhomogeneities which could reduce the scintillation light collection. Therefore, it might be that BC-404 plastic scintillator is better suited for long scatterers. For what concerns the wrapping material, if the VM2000 reflecting film is employed instead of TETRATEX, a higher tagging efficiency is found.
A scatterer characterization aimed to find best polarimetric results must be performed, there is thus the possibility to improve significantly the polarimeter sensitivity with respect to the tested set-up.

\section*{Acknowledgement}
This work was supported by a fellowship of the PhD  school in Astronomy of University of Rome ``Tor Vergata".


 \section*{References}

\bibliographystyle{elsarticle-num-names}
\bibliography{References}   

\end{document}